\documentclass[MSNbibl,dvips]{arxstspdf}
\usepackage{flushend}
\usepackage{stfloats}

\volume{26}
\issue{3}
\pubyear{2011}
\firstpage{329}
\lastpage{331}
\doi{10.1214/11-STS352REJ}
\referstodoi{10.1214/11-STS352}

\begin{document}
\begin{frontmatter}

\title{Rejoinder}
\runtitle{Rejoinder}
\pdftitle{Rejoinder}

\begin{aug}
\author[a]{\fnms{D. A. S.} \snm{Fraser}\ead[label=e1]{dfraser@utstat.toronto.edu}}
\runauthor{D. A. S. Fraser}

\affiliation{University of Toronto}

\address[a]{D. A. S. Fraser is Professor,
Department of Statistics,
University of Toronto, Toronto,
Ontario, Canada M5S 3G3 \printead{e1}.}
\end{aug}



\end{frontmatter}

\section{Introduction}

Very sincere thanks to the discussants for choosing to enter a virtual
minefield of disagreement in the development history of statistics.
For we just need to recall the remark that fiducial is Fisher's
``biggest blunder'' and place it alongside the fact that fiducial was
the initial step toward confidence, which arguably is the most
substantive ingredient in modern model-based theory: the two differ in
minor developmental detail, with fiducial offering a probability
distribution as does Bayes and with confidence
offering just probabilities for intervals and special regions.
Statistics has spent far more time attacking incremental steps than it
has seeking insightful resolutions.

As a modern discipline statistics has inherited two prominent
approaches to the analysis of models with data; of course such is not
all of statistics but is a~critical portion that influences the
discipline widely.
How can a discipline, central to science and to critical thinking, have
two methodologies, two logics, two approaches that frequently give
substantially different answers to the same problems. Any astute person
from outside would say, ``Why don't they put their house in order?'' And
any serious mathematician would surely ask how you could use a lemma
with one premise missing by making up an ingredient and thinking that
the conclusions of the lemma were still available. Of course, the two
approaches have been around since 1763 and 1930 with regular
disagreement and yet no sense of urgency to clarify the conflicts. And
now even a tired discipline can just ask, ``Who wants to face those old
questions?'': a fully understandable reaction! But is complacency in the
face of contradiction acceptable for a central discipline of science?

A statistical model differs from a deterministic model in having added
probability structure that describes the variability typically present
in most applications. So, in an application with a statistical model
and related data it would then seem quite natural that that variability
would enter the conclusions concerning the unknowns in an application:
what do I know deterministically, and what do I know probabilistically?

And that is what Bayes proposed in 1763: probability statements
concerning the unknowns of an investigation. Many have had doubts and
said there was no merit in the proposal; and many have acceded and
became strong believers.
And then Fisher (\citeyear{Fi30}) also offered probabilities concerning the
unknowns of an investigation, but by a different argument, and the turf
fight began! Bayes had hesitantly
examined a special problem and \textit{added} a random generator for the
unknown parameter, and Fisher had worked more generally and used just
the randomness that had generated the data itself.

But then a third person, Lindley (\citeyear{Li57}), from the same country said
that the second person, Fisher, couldn't use the term probability for
the unknowns in an investigation, as the term was already taken by the
first person, Bayes. And strangely the discipline complied! Decades
went by and anecdotes were traded and things were often vitriolic.

\section{What Does the Oracle Say?}

Consider some regular statistical model $f(y;\theta)$, together with a
lower $\beta$-confidence bound $\hat\theta_\beta(y)$, and also a lower
$\beta$-posterior bound $\tilde\theta(y)$ based on a prior $\pi(\theta
)$: What does the oracle see concerning the usage of these bounds?
He can investigate any long sequence of usages of the model, and He
would have available the data values $y_i$ and of course the preceding
parameter values $\theta_i$ that produced the $y_i$ values; He would
thus have access to $\{(\theta_i,y_i)\dvtx i=1,2,\ldots\}$.

First consider the lower confidence bound. The oracle knows whether or
not the $\theta_i$ is in the confidence interval $(\hat\theta_\beta
(y_i), \infty)$, and He can examine the long-run proportion of true
statements among the assertions that~$\theta_i$ is in the confidence
interval $(\hat\theta_\beta(y_i), \infty)$, and He can see whether
the\vadjust{\goodbreak}
confidence claim of a~$\beta$-proportion true is correct. In agreement
with the mathematics of confidence, that proportion is just~$\beta$.

Now consider the lower posterior bound. The oracle knows whether $\theta
_i$ is in the posterior interval $(\tilde\theta_\beta(y_i), \infty)$,
and He can examine the long-run proportion of true statements that
$\theta_i$ is in the posterior interval $(\tilde\theta_\beta(y_i),
\infty)$. Now suppose the long-run pattern of~$\theta_i$ values just
happened to correspond to the pattern $\pi(\theta)$; then in full
agreement with the mathematics of the Bayes calculation, the Oracle
would see that long-run proportion of true statements among assertions
that~$\theta_i$ is in the posterior interval $(\tilde\theta_\beta(y_i),
\infty)$ was correct, was just the stated~$\beta$.

But what if the long-run pattern of $\theta_i$ values was different
from the introduced $\pi(\theta)$ pattern? Then in wide generality the
long-run proportion of true~sta\-tements among assertions that $\theta_i$
is in the posterior interval $(\tilde\theta_\beta(y_i), \infty)$ would
\textit{not} be $\beta$! In other words, the confidence procedure is
always right, and the Bayes procedure is typically \textit{wrong}, unless
the prior was guessed correctly. Seems like a poor trade-off!

Now consider further what a prior actually does in producing parameter
bounds or quantiles that are different from the confidence bound. From
an\vspace*{1pt} asymptotic viewpoint a prior can be expanded as $\exp( a\theta
/n^{1/2}+c\theta^2/n)$ to the third order, as mentioned but not pursued
in Section~6(iv). This provides a direct displacement of the confidence bound
in standardized units and produces an $O(1)$-shift away from the
claimed $\beta$ value, either up or down depending on the sign of $a$!
Hardly an argument for using the Bayes procedure unless there was some
very urgent need for a quick and dirty calculation.

\section{Response to the Discussants}

\subsection*{Christian Robert}

Christian presents a very committed Bayes viewpoint and quite correctly
admonishes me for not distinguishing what Thomas Bayes did and what has
\mbox{followed} in the same theme. But going beyond the minor detail,
Bayes \textit{added} a distribution for a parameter, a distribution that
was not part of the binomial example under consideration and then used
that distribution for probability analysis. And much of modern Bayesian
statistics does precisely that: introduces an artifact distribution for
expediency or convenience and then works reassuringly within accepted
probability calculus. Indeed, this is the primary theme of the article:
adding something arbitrary gives something arbitrary no matter how
attractive the material labeled probability might or might not be, or
no matter what might be available\vadjust{\goodbreak} by other methods of analysis. If one
faces a~probabi\-lity-type claim, it is fair enough to simulate and
evaluate the claim, and that is what coverage probability is all about,
as the invincible Oracle well knows.

The marginalization paradoxes do appear in the literature but are
widely neglected and not ``extensively discussed'' as
Christian suggests. They apply to any proposal for a distribution to
describe an unknown vector parameter, whether obtained by the Bayes
inversion of a density or the frequentist inversion of a pivot, such as
fiducial, confidence structural or other. There is an immutable
contradiction built into the hope to describe a vector parameter by a
distribution. Curvature of an interest parameter has emerged as the
critical source for this contradiction. Take a bivariate parameter, a
data point and an interest parameter value: if the parameter is linear,
the confidence and the Bayes values are equal; if then parameter
curvature is introduced, we have that the confidence value and the
Bayes value change in \textit{opposite} directions! One has the coverage
property and the ``other'' acquires bias at twice the rate of the
departure from linearity. And the ``other'' uses the name probability
with an assertiveness coming from the use of the probability calculus,
conveniently overlooking that an artifact was introduced in place of
the input needed for the validity of the probability calculus for the
application.

Maybe it is time to address the Pandora's box and check for a Madoff
pyramid: too good to be true.

\subsection*{Larry Wasserman}

Larry presents a pragmatic view of the Bayes approach, acknowledging
its rich flexibility but recommending coverage cautions. His five
examples are most welcome concerning the wider spheres of application
and he is to be complemented
on the skillful innovations. I do quarrel, however, with his
reinforcement of personality cults in statistics. It seems that
statistics has suffered greatly from this externalization of the
scientific method, as if there were different flavors of scientific
thinking and mathematical logic and that these might gain concreteness
when personalized.

\subsection*{Kesar Singh and Minge Xie}

Confidence for estimation and exploration? It is deeply unfortunate
that statistics chooses at many steps to malign its major innovators,
for example, Fisher with his ``biggest blunder'' as a referent for the
initiative that gave us confidence. What Fisher didn't do was present
his major innovations in a fully packaged form ready to withstand a few\vadjust{\goodbreak}
centuries of challenges and modification: What? We still have to do a
little bit of thinking! Tough! He clearly must generously have expected
others to have his insight and wisdom!

Fiducial, confidence, structural or other? It is just pivot inversion
with variation in context, conditions or interpretation: the big risk
was described by Da\-wid, Stone and Zidek (\citeyear{Da73}) and curvature is now
identified as the prime cause. To have different na\-mes to fine tune for
different applications or different explorations would seem to take
emphasis away from the proper calibration of the tool, as the primary
concern for most applications.

Statistics routinely combines likelihoods as appropriate, so it is not
correct to attribute this to Baye\-sian learning; perhaps the central
sectors of statistics were just slow to glamorize the good things in
their statistical modeling. Putting a prior on a likelihood is a
different operation downstream from assembling the likelihood in the
relevant broader context, although it does seem convenient for the
Bayes approach to co-opt it as their own contribution when it was
somewhat neglected by the ``others.''

\subsection*{Tong Zhang}

Where does the pivot come from? Fisher's development of confidence or
whatever attracted the mathematicians' criticisms, mostly because it
wasn't proposed in a fully developed form. It was then shredded, fully
ignoring the emerging recognition of its innovative genius. Certainly
the need to clarify the origin of the key ingredient, ``the pivot,'' is
of fundamental importance, as Tong suggests: using all the data in an
appropriately balanced way, respecting continuity and parameter
direction from data, and more. Whether these should be bundled under a
term optimality may be questionable, but doesn't diminish the
importance of the individual criteria; for some recent emphasis on
continuity see Fraser, Fraser and Staicu (\citeyear{Fr10}).

\section{Some Concluding Invocative Remarks}

An inference distribution for a vector parameter is inherently a
contradiction. Information from two different sources can be reported
separately, with combination \textit{not} by principle.\ Combining
likelihoods is a consequence of combining models, typically following
from independence; the Bayes claim that it comes from the use of the
Bayes argument is after the fact and disingenuous.
Inverting a density and inverting a pivot are different except in the
linear case, but the first can sometimes approximate the second.

The question was asked: ``Is Bayes posterior just quick and dirty
confidence?'' And the case was made for ``Yes'': Bayes posterior is just
quick and dirty confidence: quick in the sense of easier than using
quantiles to determine how $\theta$ affects data; and approximate in
the sense of a wide spread need to use approximation methods.

Not everyone liked the blunt question. One discussion expresses
discomfort with such a direct confrontation to the Bayes approach; one
discussion adds additional support examples; and the two remaining
discussions speak more to methods and modifications of confidence
distributions, overlooking the risks. But no one argued that the use of
the conditional probability lemma with an imaginary input had powers
beyond confidence, supernatural powers.

\vspace*{3pt}\section*{Acknowledgment}
The author acknowledges the support of the Natural Sciences and
Engineering Research Council of Canada.

\end{document}